# Inventorship in AI-Assisted Inventions: Designing an Experiment to Shape Case Law

*Yevhenii Shchetynin [Corresponding Author], University of Turin, Turin, Italy, evgeniy.schetinin@gmail.com*

*Duygu Usta, University of Turin, Turin, Italy, av.duyguusta@gmail.com*

*Bryan Khan, University of Turin, Turin, Italy, bkhan.mail@gmail.com*

**Abstract**

The latest improvements in artificial intelligence (AI) raise new challenges for intellectual property laws, particularly concerning the inventorship issue in AI-assisted inventions — that is, those in which AI is used in the inventive process. While most jurisdictions allow only a natural person to be considered the inventor, the question of how to deal with AI-assisted inventions remains relevant. Namely, what is the nature and contribution of AI tools in an AI-assisted invention that would prevent a human from being recognized as its inventor? The main challenge in addressing this question is the lack of case law on the issue.

It is reasonable to assume that with the development of AI and the growing interest in its use in the inventive process, new cases will naturally arise, which in turn will harmonize and address the inventorship issue in AI-assisted inventions to some extent. However, this process will take significant time and may not keep pace with the rapid development of AI, nor fully address the new problems that arise alongside AI advancements.

This research proposes the conditions of an experiment to create relevant case law. This experiment could be initiated by society, involving stakeholders specializing in AI. The article also proposes a methodology for conducting the experiment and selecting cases that best reflect the current state of AI use in the inventive process. Conducting such an approach will help identify the most effective methods for measuring human contribution to AI-assisted inventions when determining inventorship.

**Acknowledgments**

We extend our heartfelt appreciation to the individuals, including classmates and friends, whose contributions and support were essential for completing this research. We are deeply grateful for the guidance and insightful feedback provided by Luciano Bosotti, our research advisor; and Professor Alessandro Enrico Cogo, Scientific Director of the Turin LLM-IP program, during which this research was initiated.

We also express our gratitude to the University of Turin (UNITO), COREP, and the World Intellectual Property Organization (WIPO) for facilitating our LL.M. program. Finally, we appreciate the expertise and insights shared by Associate Professor Alexey Zaytsev from BIMSA in the field of Machine Learning.

**Statements and Declarations**

The authors have no relevant financial or non-financial interests to disclose. No funding was received for conducting this study.

## 1. Introduction

Most artificial Intelligence (AI) systems use machine learning (ML) methods, such as artificial neural networks, k-means method, genetic algorithms, and generative AI, including large language models (LLMs). The main tasks AI aims to solve include classification, clustering, regression, anomaly detection, and others. Fundamental AI models have been under development since the 1940s. However, due to limitations in computing power and data storage, the accuracy of such models only began to approach human-level performance after the 2010s (Kiela, et al., 2021). From the early 2020s, generative AI, including LLMs, started to develop significantly. These technologies enable the production of new content such as text, images, sounds, and more. Recent studies also identified that machine invention systems based on AI are even capable of assisting in creating inventions (Vasilescu & Filzmoser, 2021). More recent studies (Pipada, et al., 2026) has shown rapid transformation and acceleration of drug discovery and development using AI technologies. In response to the growing use of AI, the U.S. Food & Drug Administration (FDA) has published guidelines addressing the use of AI and promoting good AI practices in drug development (FDA, 2025), (FDA, 2026).

Therefore, these innovations and improvements in AI raise new challenges for intellectual property (IP) laws, particularly concerning the issue of inventorship in inventions where AI (or any other computer tool) is used in the *inventive (or creative/invention) process* (hereinafter referred to as AI-assisted inventions). Thus, the inventorship issue in AI-assisted inventions can be formulated as follows: under what conditions can AI be assumed to have contributed to the invention, and if so, *what is the nature (and contribution) of AI tools in an AI-assisted invention such that a human cannot be considered as its inventor? Or, in other words, what level of human contribution to an AI-assisted invention is considered necessary and sufficient to be recognized as its inventor?*

The relevance of this issue arises from numerous initiatives from patent offices in different countries aimed at developing guidance to address it. For example, the Japan Patent Office (JPO) recently pointed out AI is expected to play a broader role in the inventive processes, see (METI, 2024). Similar issues also arose in academic literature. Recently The AI Scientist (The AI Scientist, 2025a) system based on LLM models introduced probably the first fully AI-generated scientific paper that has passed the peer-review process on ICLR 2025 Workshop (The AI Scientist, 2025b).

The inventorship issue in AI-assisted inventions was also recently addressed by the United States Patent and Trademark Office (USPTO) in their "Inventorship Guidance for AI-Assisted Inventions" (USPTO, 2024a). The USPTO introduced five basic principles based on the *Pannu* factors (Pannu v. Iolab Corp., 1998) and provided two artificial examples (one is related to the use of a LLM model and another with the use *in silico* methods in drug discovery), see (USPTO, 2024b). These examples illustrate how the principles can be applied to measure human contribution. At the same time, they admitted that there is "*no bright-line test*" to determine human contribution to AI-assisted inventions, indicating that the question requires further development due to the lack of case law and practical application of the proposed principles.

Although, in November 2025, the USPTO published revised inventorship guidance (USPTO, 2025a) rescinding previous guidance, they still highlight that only natural persons can be named as inventors, while AI is referred as a tool in hands of humans. The previous approach based on Pannu factors to AI-assisted inventions was withdrawn with arguments that those factors only apply for joint inventorship between natural persons and are inapplicable to AI. Regarding AI-assisted inventions USPTO referred to the case law (Shatterproof Glass Corp. v. Libby-Owens Ford Co., 1985) and which (Hess v. Advanced Cardiovascular Sys., 1997) which establishes that inventors may "*use the services, ideas, and aid of others*" with conclusion that the same principle should apply to AI systems.

However, such strong interest from patent offices together with rapid changes in legislation, as well as the pace of AI development, which may lead to another revision of laws or guidelines, only underscore the relevance of the topic arises in this study: under what conditions can AI be assumed to have contributed to an invention, and if so, what level of human contribution to an AI-assisted invention is considered necessary and sufficient to be recognized as its inventor, and how might examination procedures be adapted?

Since most jurisdictions currently allow only natural persons to be considered inventors, answering the question posed will help applicants determine in advance which inventions are worth patenting and which might be better protected by other methods (such as trade secrets, publication etc.), at the same time reducing the workload of the patent offices. On the other hand, if AI were to be recognized as an inventor in the possible future, as proposed by some authors (Mohamed, Rahma, & Hemairy, 2025), addressing this question would help us adapt more quickly to these new realities.

One of the problems with addressing this question is the lack of case law on the issue, see (Nemec & Rann, 2023). Some of the well-known cases are the so-called DABUS cases. These involve a food container created by DABUS, an AI tool developed by Stephen L. Thaler (as part of his "Imagination Engines Incorporated" project (Thaler S. , 2023)), who listed DABUS as the sole inventor and applied for patents in several jurisdictions. Most jurisdictions rejected the patent applications on the grounds that only a natural person can be indicated as an inventor (e.g., in the United States of America and China). It should be noted that most courts did not address the question of joint inventorship (also called co-inventorship) between humans and AI tools, nor whether an invention should be granted if Thaler were to amend the application to list himself as the inventor.

It is reasonable to assume that, with the development of AI and the growing interest in its use in the inventive process, new cases will naturally arise, which in turn will harmonize and address the inventorship issue in AI-assisted inventions to some extent. However, this process will take significant time and may not keep pace with the rapid development of AI or address new problems arising alongside AI advancements.

Moreover, there already exist AI-systems which help patent attorneys to autonomously answer for office actions from patent offices (Chu, Lo, Hsiang, & Cho, 2024). Another study (Lee, 2024) introduces AI system based on LLM models to help draft patent claims that may increase the probability for a patent of being granted. It is obvious to assume that such AI systems can be used to suggest amendments into invention claims and thus into conception of the idea, although it should be made in consistence with description submitted by applicant.

Therefore, this research discusses the conditions under which a bold new experiment could be implemented to create relevant case law (or presidential rulings/determinations at the administrative level of the national patent office (which are not technically "cases")). Additionally, a methodology is proposed for conducting this experiment and selecting cases that best reflect the current state of AI use in the inventive process. This experiment could be initiated by society, involving all stakeholders — companies using or developing AI, academic institutions, government entities including legislators and patent offices, and individuals. Implementing this approach will help identify effective methods for assessing human contribution to AI-assisted inventions in determining inventorship. Engaging diverse participants would also make it possible to compare whether the perceptions of law and patent offices align with those of industry and other stakeholders.

## 2. State of the art

### *2.1. Traditional Notion of Inventorship in Patents*

Inventorship refers to the individual or individuals who have made significant and original contribution to the invention that is the subject of the patent. This may be a contribution to the conception of the idea or the subject matter of at least one claim included in the patent application. Inventorship provides the starting point for determining ownership of patent rights (Chisum, 1978).

Despite ongoing discussions regarding the notion of inventorship, particularly with the rise of AI, the traditional patent system recognizes only natural persons as inventors, since inventorship requires a mental act of creativity exclusive to humans. This principle aligns with the historical foundations of intellectual property, which naturally emphasize the necessity of human element (Saw & Chan, 2023). Intellectual property is grounded in three main theories: the labour theory, which asserts that individuals should benefit from their productive labour (Hull, 2009); the personality theory, which views person's creations as an expression, i.e. an extensions of an individual's identity (Radin, 1982); and the utilitarian theory, which justifies rewards for contributions that benefit society (Mill, 2004). In accordance with these theories that are the basis of IP, mandating that inventorship be exclusively assigned to natural persons guarantees that the rights associated with patents are directly linked to the people whose creative efforts contribute to the invention.

In most jurisdictions, there is no legal obligation for patent applicants to disclose the process by which their invention was created. Consequently, determining the true contributors to an invention can pose significant challenges. This lack of transparency in the patent application process not only challenges attributing inventorship but also may raise concerns regarding the integrity and fairness of the patent system.

In most jurisdictions, the designation of inventors is declaratory in nature — the inventors are the individuals named in the patent application, unless proven otherwise. If an individual believes they deserve recognition as a co-inventor but are not listed on the patent application, they may seek legal recourse through litigation or administrative review processes. This mechanism is intended to incentivize applicants to disclose all legitimate co-inventors correctly. A similar situation arises when the list of inventors includes individuals who have not made a creative contribution and should be removed from this list. An already issued patent may even be invalidated if the list of inventors turns out to be incorrect. However, this logic breaks down in the case of AI, which cannot assert legal rights or initiate claims, thereby undermining the effectiveness of administrative oversight.

But at the same time, this gap in Patent Law creates legal uncertainties about inventorship issue for applicants. If applicants will not designate all human inventors, or if they designate human inventors who haven't made any inventive contribution in an invention fully made by AI, in the future such an invention could be invalidated in court if someone (e.g. competitor) finds this discrepancy in the list of inventors.

### 2.2. *Criteria for identifying and challenging inventorship*

The criteria for identifying and challenging inventorship vary significantly across jurisdictions, reflecting diverse legal traditions and policy considerations. This section examines the criteria and processes in three members of the IP5[1] — USPTO, the European Patent Office (EPO, representing Europe), and the JPO — each with advanced patent systems, strong economies, and representing both common law and civil law traditions. These jurisdictions have already begun addressing inventorship issues arising from AI-assisted inventions — with varying degrees of progress — making them particularly relevant case studies. While most other countries operate under broadly similar patent law principles, the IP5 examples provide a clear and influential benchmark for global practice.

**The United States of America (USA)**

Under 35 U.S.C. § 116 (USPTO, 2022), the United States Patent and Trademark Office sets forth the requirements for inventorship. According to this guidance, an inventor is defined as an individual or a group of individuals who contribute to the conception of the claimed invention. This means that inventorship is determined based on the patent claims rather than the specification or drawings. Each inventor must have made a contribution to at least one claim of the patent, and this contribution must be included in the final set of claims as granted.

Challenges to inventorship in the U.S. can arise if it is discovered that an inventor listed did not contribute to the conception of the claimed invention or if an unlisted individual did. Such disputes often require legal proceedings to resolve, with substantial evidence needed to prove or disprove the contributions in question. The USPTO may require affidavits or other documentation to support claims of inventorship or to contest them.

**Europe**

The European Patent Office mandates that every patent application must designate the inventor, who must be a natural person. The EPO guidelines (EPO, 2024) specify that the EPO does not verify the accuracy of the information provided in the designation of the inventor but ensures that the designated inventor is a natural person, as stipulated in case law (J 0008/20 (Designation of inventor/DABUS), 2021).

If the designation of the inventor is not filed, or if it cannot be considered valid due to deficiencies (such as missing name, country, place of residence, or applicant's signature), the EPO will inform the applicant that the European patent application will be refused if the deficiency is not remedied within the prescribed period under Rule 60(1). This strict adherence to formal requirements ensures that the inventor's designation is complete and accurate, preventing potential legal issues related to inventorship.

Challenging inventorship at the EPO typically involves addressing these formal requirements. If there is a dispute over who should be named as an inventor, the challenge must be resolved through legal or administrative proceedings, with the EPO ensuring compliance with procedural rules rather than evaluating the substantive contribution of each inventor.

**Japan**

According to the guidelines provided by the Japan Patent Office, see (JPO, 2015), the criteria for recognizing inventorship are based on the contributions made to the invention as described in the specification and drawings. Unlike some other jurisdictions including the USA, Japan does not determine inventorship solely based on patent claims. This means that any individual who has contributed to the inventive concepts detailed in the specification and drawings must be listed as an inventor. This approach ensures that all contributors to the inventive process are duly acknowledged.

Challenging inventorship in Japan can involve proving that the listed inventors did not actually contribute to the invention as described or, conversely, that someone not listed made a significant contribution. The JPO's emphasis on the specification and drawings as the basis for inventorship highlights the importance of comprehensive and detailed documentation during the invention process.

In conclusion, IP5 jurisdictions have different criteria for identifying and challenging inventorship, reflecting their distinct legal traditions and policy concerns. In the USA, the USPTO considers a person an inventor if they have contributed to at least one patent claim, with disputes often requiring legal action and substantial evidence. Europe, through the EPO, enforces strong formal standards for inventor designation, focusing on procedural compliance rather than substantive evaluation, and mandates that inventors must be natural persons. In contrast, the JPO in Japan bases inventorship on contributions detailed in the specification and drawings, ensuring all contributors are acknowledged.

[1] The five IP offices (IP5) is the name given to a forum of the five largest intellectual property offices in the world that was set up to improve the efficiency of the examination process for patents worldwide: USPTO, EPO, JPO, the Korean Intellectual Property Office (KIPO), the National Intellectual Property Administration of the People's Republic of China (CNIPA).

### *2.3. Joint inventorship*

Joint inventorship occurs when two or more individuals collaboratively contribute to the conception of an invention. Unlike sole inventorship, where a single individual is responsible for the invention, joint inventorship recognizes the collaborative nature of the inventive process. This collaboration can involve a diverse array of contributions, each of which is essential to the final invention.

In the United States, the criteria for joint inventorship are guided by Section 35 U.S.C. § 116, (USPTO, 2022) which stipulates that there is no need for inventors to work together physically or at the same time, nor do they need to contribute equally to the invention. Instead, each inventor must make a significant contribution to the conception of the invention as articulated in the patent claims.

The determination of joint inventorship in the U.S. is further clarified by the (Pannu v. Iolab Corp., 1998) case, which established several key factors known as the *Pannu factors,* which ensures that joint inventorship is not merely a formality but a recognition of genuine collaborative effort that results in the creation of an invention.

Issues related to joint inventorship are particularly pertinent in the context of AI-human contributions to inventions. Complex collaborations can arise in both scenarios, where it's hard to pinpoint the precise contributions of each party. For instance, determining the contributions of an AI system versus a human inventor raises questions is similar to determining the contributions of each inventor in joint inventorship cases:

*Attribution of Contributions:* It can be challenging to clearly attribute specific contributions to human or AI, much like distinguishing individual contributions in joint inventorship.

*Inventive Step:* Assessing whether the AI's output constitutes an inventive step and how it integrates with the human inventor's input parallels the evaluation of significant contributions among co-inventors.

Both situations demand a nuanced understanding of the inventive process and careful documentation to ensure that all contributions are accurately captured and credited. Therefore, managing joint inventorship and joint ownership requires careful legal and contractual planning to mitigate potential disputes and ensure that the rights and contributions of all parties are fairly recognized and protected.

Before exploring the complexities of AI-assisted inventorship, it is crucial to understand the traditional definition of inventorship in patents. Inventorship refers to individuals who have made significant and original contributions to the conception of an invention, a concept rooted in the belief that only natural persons can perform the necessary creative or inventive mental acts. This principle, based on theories of labor, personality, and utilitarianism, ensures that patent rights are connected to human creators. However, there are notable regional differences in the criteria for determining and challenging inventorship. As discussed above, in the United States, inventorship is based on contributions to patent claims, often resolved through legal processes. Europe, through the EPO, enforces strict formal standards focused on procedural compliance. In Japan, the JPO determines inventorship based on contributions detailed in specifications and drawings.

Furthermore, as U.S. case law has established, joint inventorship recognizes collaborative contributions from several parties, requiring a substantial amount of work from each contributor even in the absence of simultaneous or physical collaboration. To ensure fair recognition and protection of each party's rights, meticulous documentation is essential because the complex nature of AI-human inventorship complicates the attribution of contributions and the evaluation of inventive step.

### *2.4. Inventorship in AI-assisted Inventions*

The notion of inventorship has started to be questioned, particularly in patents where the inventive process involves AI assistance. This is because in many legal jurisdictions, applicants are not required to disclose the process by which their invention was created. Consequently, challenges to inventorship typically arise during opposition procedures or subsequent legal proceedings. This explains the limited number of cases concerning inventorship, particularly in the context of AI-assisted inventions. This situation underscores the relevance of the proposed approach outlined in the current research, wherein it is advocated for distributing the responsibility of resolving inventorship issues in AI-assisted inventions to both patent examiners and judges. However, before proposing possible solutions, it is essential to analyze the inventions involved.

Depending on the different proportions of human and AI contributions/involvements (hereinafter human-AI contributions/involvements) in the inventive process of inventions there are three main categories:

1. Inventions fully made by human(s) without using AI tools (simply — inventions/traditional inventions/human inventions).
2. Inventions made with minor AI-assistance (AI-assisted inventions).
3. Inventions made with significant AI-assistance (so-called AI-generated inventions).

Traditional inventions (1) are solely created by humans and the inventorship is credited to the individual or individuals involved in the patent's creation. Under this framework, there is clear certainty regarding inventorship, with potential discussions focusing on the level of contribution from each person, especially in cases of joint inventorship.

Nonetheless, the significance of this distinction is minimal, as even a single claim may be enough to determine someone as an inventor.

AI-assisted inventions (2) will be discussed in the following sections in more detail. This type of invention is of the greatest interest, because it is in relation to them that this paper proposes an approach to determine human-AI contributions.

So-called AI-generated inventions (3) refer to situations where AI contributes to the invention process but don't independently conceptualize the inventive idea. AI acts as a tool or aid to human inventors. These include human-made inventions using AI for outcome verification and inventions where a human identifies a problem and employs AI to find a solution. It also includes the (largely hypothetical) scenario where inventions are made purely "autonomously" by AI systems. At the moment, patent protection is not provided for this type of invention in most jurisdictions. This was resolved in DABUS cases that will be discussed in more detail below. Although in certain situations a dependent claim that recites a minor contribution made entirely by AI may still be included in an application, so long as the underlying independent claim has a valid human inventor who made a significant contribution to the conception of the overall inventive concept, see (Crawford & Bockman, 2024). It is also worth noting that, despite the fact that AI-generated inventions are not protectable in general at the moment, the approach proposed in our work can be used in the future if protection is granted to AI-generated inventions (either in terms of inventorship, ownership, or sui generis rights).

### *2.5. Significant AI-assistance in inventions (AI-generated inventions)*

AI systems can provide substantial assistance in various stages of the inventive process, including data analysis, simulation and modeling, idea generation and optimization. However, in the context of AI-assisted inventions, defining what constitutes a "significant" contribution is crucial for determining inventorship.

For this purpose, case laws will be reviewed to determine which instances constitute a "significant" contribution by AI. Additionally, the examination will cover the countries where AI makes a substantial contribution but is not recognized as an inventor due to the minimal involvement of human contribution.

#### The United State of America

The United States has established a new approach to determining inventorship in AI-assisted inventions. The guidance (USPTO, 2024a), as per the USPTO, focus on the contributions of natural persons involved in the inventorship process. Key principles include:

1. The natural person can be listed as an inventor if his contribution to AI-assisted invention was significant.
2. Merely identifying a problem, general goal or research plan isn't sufficient for the natural person to be listed as an inventor. At the same time "*a significant contribution could be shown by the way the person constructs the prompt in view of a specific problem to elicit a particular solution from the AI system*".
3. Mere recognition and appreciation of the output of an AI system as an invention is not sufficient.
4. A natural person who develops an essential building block from which the claimed invention is derived may be considered to have provided a significant contribution.
5. A person simply owning or overseeing an AI system that is used in the creation of an invention, without providing a significant contribution to the conception of the invention, does not make that person an inventor.

#### Thaler v. Vidal

In the landmark Thaler case (Thaler v. Vidal, 2022), the Federal Circuit reaffirmed that inventors must be natural persons. Thaler had listed his AI system, DABUS, as the inventor for two patents. The court held that the Patent Act requires inventors to be human, thus rejecting the AI's designation as an inventor. The USPTO's decisions explained that under current U.S. patent laws, inventorship is limited to a natural person(s). *"The position expressed by the Federal Court in Thaler, that an inventor must be a natural person, and by extension, any joint inventor must be a natural person. As such, patent applications that name a machine on an application data sheet (37 CFR 1.76), an inventor's oath or declaration (37 CFR 1.63), or a substitute statement (37 CFR 1.64) as either an inventor or joint inventor will be considered by the USPTO to have improper inventorship"* (USPTO, 2024a).

Additionally, applications and patents must not list any entity that is not a natural person as an inventor or joint inventor, even if an AI system may have been instrumental in the creation of the claimed invention (USPTO, 2024a).

#### Pannu v. Iolab Corporation

This case provides a critical precedent for understanding the requirements of significant contribution. The court emphasized that an inventor must be involved in the conception of the invention, reinforcing the notion that mere supervision or ownership is inadequate for inventorship as explained in Chapter 1.

Although the *Pannu* factors are generally applied to two or more people who create an invention (*i.e.,* joint inventors), it follows that a single person who uses an AI system to create an invention is also required to make a significant contribution to the invention, according to the *Pannu* factors, to be considered a proper inventor (USPTO, 2024a). Moreover, a named inventor doesn't need to contribute to every claim in a patent application; contributing to just one claim is enough. However, each claim must be invented by at least one named inventor in contrast with Japan. In essence, a natural person must significantly contribute to each claim in a patent application or patent.

Inventorship is improper in any patent or patent application that includes a claim in which at least one natural person did not significantly contribute to the claimed invention, even if the application or patent includes other claims invented by at least one natural person. Therefore, a rejection under 35 U.S.C. 101 and 115 should be made for each claim (USPTO, 2024a).

**Europe**

The EPO has similar guidelines to the USPTO but with its unique rules. From an inventorship standpoint, AI inventions can be categorized into three types:

1. Human-made inventions use AI for outcome verification.
2. Inventions where a human identifies a problem and employs AI to find a solution.
3. AI-made inventions, where AI identifies a problem and proposes a solution without human intervention.

In the first two categories, AI is used as a tool for human inventors, augmenting their capabilities and it's a minor contribution explained in the previous chapter. In the third category (AI-made inventions), scientists seem to agree that AI which could invent independently of human direction, instruction and oversight is a matter of undefined future and thus science fiction (EPO, 2023).

There is a consensus that an inventor is a human being who creates an invention through their own creative efforts. This view is supported by an academic study on AI inventorship commissioned by the EPO and discussions with EPC contracting states (Shemtov, 2019). Moreover, the EPO mandates that the designated inventor in a patent application must be a human being, not a machine. This requirement has legal implications, ensuring that the designated inventor is legitimate and can benefit from the rights associated with this status. To exercise these rights, the inventor must possess legal personality, which AI systems or machines lack.

The legal requirement that an inventor must be a human was challenged when two applications named an AI system (DABUS) as the inventor were filed with various patent offices worldwide. In 2019, the EPO rejected these applications (see (EP Patent No. 18275163.6 (patent application), 2018), (EP Patent No. 18275174.3 (patent application), 2018)) because the EPC requires inventors to be natural persons. Appeals were filed by the applicant but were dismissed by the EPO Legal Board of Appeal during oral proceedings on December 21, 2021 (see (J 0008/20 (Designation of inventor/DABUS), 2021) and (J 0009/20 (Designation of inventor/DABUS II), 2021)). The Legal Board affirmed that under the EPC, the inventor must have legal capacity, and a statement indicating the origin of the right to the European patent must specify the inventor's successor in title. On December 20, 2021, the applicant filed divisional application (EP Patent No. 21216024.6 (patent application), 2018), which is currently under examination.

**Germany**

German Patent Law, while referencing *"inventorship"* in Sections 6-8, 37, and 63, lacks a clear definition of what constitutes an *"invention"*. German case law stipulates two conditions for being considered an inventor: (1) the person must make a substantial, creative contribution to solving the problem addressed by the invention, although this contribution need not be inventive; and (2) the contribution must be intellectual — i.e. it must originate from the person, and not from following instructions by others. This means the contribution must be significant to the invention's success and result from intellectual effort (Shemtov, 2019).

In the recent case (Thaler v. German Patent and Trade Mark Office, 2024), the Federal Court of Justice of Germany ruled that AI-generated inventions are patentable and that a natural person must be designated as the inventor, even if AI played a role in developing the technical teaching. *The decision states that providing a sufficiently clear designation of the inventor, such as indicating that the inventor directed a specific artificial intelligence to generate the invention, is legally irrelevant and does not justify the rejection of the application.*

The criteria established by the Court for obtaining a patent on an AI-generated invention are as follows:

- A natural person can be designated as the inventor even if AI was used to develop the claimed technical teaching.
- A human contribution that significantly influenced the success of the invention is sufficient for the person to be considered the inventor, even if AI assisted in the discovery.
- The specific type or intensity of the human contribution needed for inventorship is debated, but not crucial to the determination.
- It is still possible to identify human contributions when using AI, and legal assessment can establish inventorship based on this, as no AI system operates entirely without human input.

- Even if AI made a substantial contribution, there should be at least one natural person recognized as the inventor based on their contribution, making it reasonable to name a human inventor.

In (Kim, 2025) author argues that the German Federal Court of Justice's decision in the DABUS case adopted a surface-level, formalistic approach to inventorship in AI applications, effectively reducing inventor designation to a legal fiction by accepting "any human" as sufficient while sidestepping the substantive question of meaningful human contribution.

Thus, even though it is possible to obtain a patent in Germany in the case of DABUS if a natural person is designated as an inventor, the formality of such a decision leaves the legal uncertainty regarding human contribution in AI-assisted inventions unresolved.

**The United Kingdom (UK)**

The Intellectual Property Office of the United Kingdom (UKIPO) aligns with the EPO and USPTO, requiring that inventors be natural persons.

Section 7(3) of the UK Patent Act 1977 (UKIPO, 2021) defines the inventor as *"the actual deviser of the invention"*. The Court of Appeal clarifies that *"invention"* in this context refers to the information in the specification, not just the claims, since entitlement issues can arise even before claims exist. The inventive concept, which must be identified, refers to the core idea or *"heart of the invention"* disclosed in the specification. The *"deviser"* is someone who contributes intellectually to the inventive concept, beyond financial, abstract, or administrative input. This contribution must involve aspects not disclosed in prior art and potentially rendering the invention patentable ("*contribution must be to the formulation of the inventive concept"* — see (Yeda Research and Development Co Ltd v Rhone-Poulenc Rorer International Holdings Inc House, 2007)).

The "*deviser*" typically means a human actor, as AI lacks the thought process required for devising an inve*ntion. AI can be crucial in developing an invention but is seen as a tool rather than an inventor. The human who utilizes AI to contribute to the inventive concept is considered the inventor.*

In view of DABUS, UKIPO made a consultation with society (UKIPO, 2021a) regarding how patent system should deal with AI technology and particularly about protection of AI-devised (AI-assisted) inventions. UKIPO received 48 responses (see (UKIPO, 2021b) and (UKIPO, 2022)) about the view of patents and AI inventorship and concluded that UKIPO will not change current patent laws, as existing rules are seen as sufficient for protecting AI-assisted inventions. The UK government highlighted that they would pursue international discussions on AI inventorship to support UK innovation and competitiveness, while continuing to assess future IP needs for AI-assisted inventions. Also, from the report it can be concluded that they generally include generated AI-generated inventions in the set of AI-assisted inventions and thus AI-generated inventions might receive patent protection if application satisfies the legal requirements, including novelty and non-obviousness and human-inventorship. Thus still, in certain cases AI-generated inventions might not satisfy non-obviousness criteria if human contribution wasn't inventive.

**South Africa**

In South Africa, the AI system DABUS was acknowledged as an inventor for a patent. This marked the first time an AI system was granted patent rights as an inventor, signaling a notable shift in the field.

However, unlike some other patent offices, South Africa's patent office does not conduct substantive examinations of patent applications. Instead, it focuses on ensuring that applications meet the necessary formal requirements. This means that no substantive decision was made regarding the invention's merit or the suitability of an AI system as an inventor within this process. The patent office did not formally assess the legitimacy of the AI as an inventor but primarily ensured the application forms were complete and correct, see (Inventa, 2023).

**Canada**

The recent decision (Thaler S. L., 2025) of the Commissioner of the Canadian Patent Office and the Patent Appeal Board in the DABUS case refused the application for the food container. In this decision, the Commissioner and the Board reviewed the patent application, oral hearing arguments, and cited articles submitted by Thaler.

The main conclusions regarding inventorship were as follows:

- The term "inventor" is limited to a natural person or persons.
- The filed application cannot be patented because DABUS is an AI system. Dabus cannot be recognised as an inventor and, furthermore, the application cannot be filed by "the inventor or the inventor's legal representative," as required by the Canadian Patent Act.
- A valid inventor cannot be identified. Therefore, there is no "inventor's legal representative" who could file the application, even though it was submitted by Thaler and all required fees were paid.

The analysis of articles related to the use of AI in drug development conducted in this Decision showed that there was "*no discussion of per se inventing by AI systems*". In other words, mentioned articles describe the use of AI as a tool in the drug development process, rather than AI systems independently inventing a new drug. Thus, the Decision indirectly established the position that AI-generated inventions cannot be patented. At the same time, inventions in which AI was used merely as a tool may still be patentable, although no analysis of the sufficient amount of human contribution in such cases was conducted.

The global landscape for AI-assisted inventions is summarized in table 1 and it shows a strong consensus on the necessity of human inventorship. Currently, most jurisdictions require significant human contribution and do not accept AI as an inventor. While AI can significantly aid the inventive process, the core inventive step must stem from human ingenuity and contribution.

*Table 1 Global landscape of patentability for AI-generated inventions*

| Country / jurisdiction | Can AI be listed as an author of invention? | Can AI-generated inventions be patented? | Comments |
|---|---|---|---|
| USPTO | NO | NO | Human contribution to AI-assisted invention should be significant. |
| EPO | NO | NO | Human contribution to AI-assisted invention should be significant. |
| Germany | NO | Technically YES, if a natural person is designated as the inventor) | While a person must make a substantial contribution to solving the problem, it need not be inventive. In the DABUS case, the Court confirmed that designating a natural person in AI-generated invention as the inventor is sufficient for the application to be accepted. |
| JPO | NO | NO | Human contribution to AI-assisted invention should be significant. |
| UKIPO | NO | Technically YES | UKIPO does not distinguish AI-generated inventions into a separate category and classifies them as AI-assisted inventions. |
| South Africa | YES | YES | There is no substantive examination, only formal examination |
| Canada | NO | NO | A valid inventor cannot be identified |

### *2.6. Challenges in Determining Human Contribution to AI-assisted Inventions*

Although there have been very recent developments on this issue, see (USPTO, 2024a), the USPTO acknowledges that *"…in determining whether a natural person's contribution in AI-assisted inventions…there is no bright-line test*". Instead, the process remains detailed and involves evaluating the specific circumstances of each case.

Kathi Vidal, Under Secretary of Commerce for Intellectual Property and Director of the USPTO emphasized that the patent system is designed to incentivize and protect human ingenuity, as well as the investments required to transform ideas into marketable products and solutions. It also incentivizes the sharing of ideas to enable further innovation. Recent guidance seeks to balance granting patent protection for AI-assisted inventions while ensuring that innovation remains accessible for future advancements. This approach acknowledges AI's role in innovation while maintaining a focus on human contribution, see (USPTO, 2024a).

In Europe in testing patentable subject matter in case law such as (Aerotel Ltd v Telco Holdings Ltd, 2006) in the UK, led also to the development of the fourth-step test. The second step of this test introduced the concept of "*technical contribution*", which notes: *"What has the inventor really added to human knowledge perhaps best sums up the exercise".*

## 3. Method

In previous sections, existing cases related to inventorship issues in AI-assisted inventions were discussed. Due to the fast development of AI technologies and legal uncertainty, relevant cases currently include only so-called DABUS cases. But these are corner cases, where AI tool DABUS was used with such a contribution that notion of inventorship was not granted to Thaler, a developer of DABUS system. Moreover, the provisions of the IP laws have just begun to adapt to new realities of emerging AI technologies, and so far only in the form of guidance for examiners or comments on the existing provisions of the IP laws. Therefore, the problem of a lack of case law and legal provisions of how to deal with inventorship issue in this emerging area of AI-assisted inventions has arisen.

Thus, this section proposes for discussion a new approach of how inventorship issue in AI-assisted inventions can be resolved. Specifically, it addresses the question of how to measure the amount of AI contribution (and vice versa — human contribution) to an AI-assisted invention that will be considered necessary and sufficient for a human inventor to be recognized as its inventor.

### *3.1. Creation of case law by applicants*

One of the objectives of this paper is to propose a mechanism to create precedents that define the test or describe the principle of how to measure human-AI contributions to AI-assisted invention in determining inventorship. The current study proposes artificially create such a case law (or presidential rulings/determinations at the administrative level of the national patent office), by creating case studies by some initiative group of stakeholders. This can be initiated by stakeholders specialized in AI — companies using or developing AI, academic institutions, government entities as well as individuals. This group will apply for patents for different inventions in different jurisdictions in order to incentivize patent offices and IP authorities of countries to create case laws in a more fast and methodological manner. Below, the rationale for the economic benefits for such an initiative group in using the proposed approach will be presented. However, for simplicity, it should be assumed that such an initiative group has been found and is ready to invest funds in the proposed experiment.

This incentive can be based on a methodology that includes the following steps:

1. Creating a set of inventions (i.e., cases) with varying degrees/levels of human and AI involvement. For the purposes of the experiment, it is assumed that all created inventions satisfy the patentability requirements. Namely, the following optional actions may also be undertaken:
   a. Manually assessing the created inventions against the patentability criteria — novelty, inventive step, and industrial applicability.
   b. Selecting, from each level of human-AI contributions, the invention most likely to meet the patentability requirements.
   c. Manually improving certain inventions based on the patentability assessment. If such improvements increase human contribution, the invention would be reassigned to a higher human-involvement level.
2. Filing patent applications by stakeholders (applicants) for selected inventions in designated jurisdictions. Each application should include a description detailing how the invention was created and specifying the respective contributions of humans and AI. This will require patent examiners to evaluate compliance with human inventorship criteria for each involvement level.
3. Pursuing appeals in case of rejection, as Court of Appeal or Supreme Courts often establish influential case law in this area.
4. Deriving long-term insights — in the coming years, society is expected to gain a clearer understanding of the threshold level between human and AI contributions to the inventive process in AI-assisted inventions that is sufficient to recognize a person as an inventor and grant patent protection.

A visual representation of the described methodology is presented below, in **Fig. 1**.

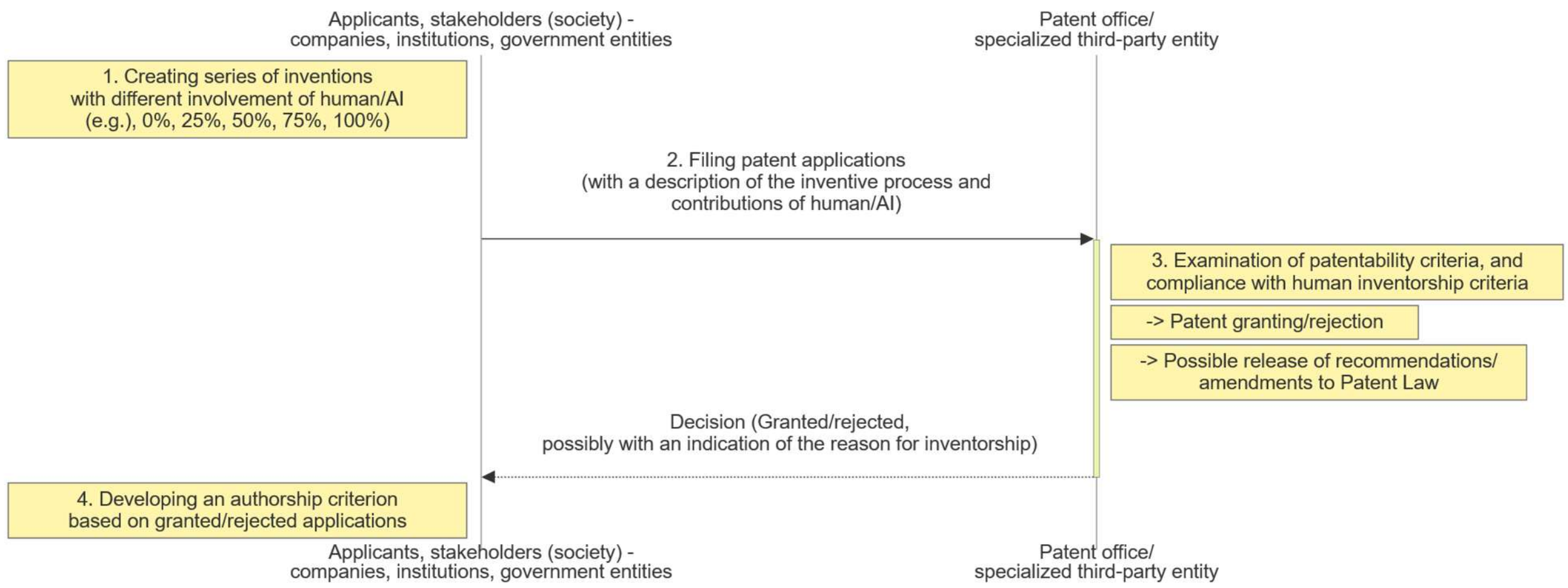

***Fig. 1*** *Methodology for creation of case law by applicants*

Regardless of personal preferences, case law will inevitably evolve for inventions involving AI in the inventive process. The approach proposed in this article will not only speed up this process, but also make it systematized, as well as more accurate in terms of developing patentability criteria. The suggestion of our research is to generate a series of inventions where AI plays a role in inventive process, alongside varying levels of human-AI contributions.

Subsequently, patent applications will be submitted to the patent office, accompanied by detailed descriptions of the inventive process. As a result of such an initiative from society, the patent office will be obliged to answer the question of what contribution of the author is sufficient to be considered an inventor. Over time, criteria for assessing the patentability of AI-assisted inventions will be refined and legal uncertainty will be reduced (legal uncertainty is a "*variance of the expected judgment of the court*" according to (Weiss, 2019)). That will stimulate technological development in the shortest possible time.

This proposal will speed up the process of creating case law. There is also an economic feasibility in this. Companies specializing in AI, as well as those utilizing AI in their R&D, may face legal uncertainty regarding the patentability of inventions involving AI tools. This legal uncertainty could lead them to prefer to have trade secrets, resulting in increased deadweight loss, undermines the function of the patent system in incentivizing socially beneficial disclosure, and slowing down technical progress.

It should be noted that step 2 of the proposed approach can also be implemented in different ways. Besides filing patent applications by stakeholders for chosen inventions in designated jurisdictions, inventions can be considered by courts, specialized third-party entity specially created to develop criteria of how to address inventorship issue in AI-assisted inventions and provide developed recommendations to patent offices. The entity which will proceed with step 2 is also a question for further research and discussion, as the main goal of current study is to provide a general methodological approach of how to resolve inventorship issue in AI-assisted inventions.

It is also worth noting that creating a set of inventions (i.e., test cases) in step 1 of the proposed approach can be done in different ways:

1. *Strict Approach.*

   In this approach, inventions (i.e. cases) are selected or created according to predefined levels that reflect varying proportions of human contribution and AI involvement (human-AI contributions) to the inventive process. Step 1 is then implemented by creating several inventions for each level using a robust methodology — for example, by organizing them within selected technology domain which can be selected using IPC (International Patent Classification) categories.

2. *Broad Approach.*

   In this approach, inventions are selected without regard to, or at a fixed value of, the human-AI contributions ratio. Instead, selection is based on (a) the type of inventive process in which AI tools were involved, and (b) the technology domain in which the patentable subject matter resides. This allows for variation in process type and technical field while controlling or disregarding contribution ratio.

A hybrid implementation of step 1 can also be employes, combining the structured classification of the strict approach with the flexibility and scope of the broad approach. However, it is worth noting that the present research only introduces a novel approach for creating case laws, thus work provides only a few possible ways of implementing it, while other methods may be developed through further research and discussion, surveys and experiments.

Mentioned implementations of the proposed approach are discussed in more detail below in the relevant subsections.

### *3.2. Strict approach in creating test cases*

To achieve better results, a separate methodology is needed for comparing inventions with different degrees/ratios of human and AI contributions. One possible way to do that is to formalize the question of how to choose case laws in a methodological way. Let's assume there are following sets: $\boldsymbol{Cases} = \{\boldsymbol{case1}, \ldots, \boldsymbol{caseN}\}$ — set of *N>1* cases, $\boldsymbol{Ratios} = [\mathbf{0}, \mathbf{1}] \subset \mathbb{R}$ — set of ratios of human contribution in AI-assisted inventions for each $case \subset \boldsymbol{Cases}$. Let's assume such a methodology is expressed in the function $\varphi$:**Cases** ↦ **Ratios**, $\varphi(case) \subset [\mathbf{0}, \mathbf{1}]$, exists, allowing comparison of ratios of human and AI contributions in AI-assisted inventions, $\varphi$ ranging from 0% to 100% based on human contribution. AI contribution is thus equal to $1 - \varphi(case)$.

For this purpose, $\varphi(case)$ must satisfy the following criteria:

1. For each two cases case1, case2 $\subset \boldsymbol{Cases}$; $ratio1\ (case1)$, $ratio2(case2) \subset \boldsymbol{Ratios}$, the following condition is met: $\varphi(case2) > \varphi(case1)$ if and only if the human contribution in *case2* is higher than in *case1*, that is ratio2 $>$ ratio1.
2. For *N* cases (*N>1*), in which human contribution varying from 0% for *case1* to 100% for *caseN*, the following condition is met: $\varphi(case1) < \varphi(case2) < \cdots < \varphi(caseN)$.
3. Assuming that the patent office's response is rational in terms of granting or rejecting patent applications due to the lack of human contribution in the inventive process and for each case satisfy the conditions of patentability, patent offices will reject *case1* with no human contribution and accept *caseN* with pure human contribution. Moreover, there exists a value $M \subset (1, N]$ such that all cases from *1* to *M-1* will be rejected, while all cases from *M* to *N* will be accepted.

A simplified example of the proposed strict approach is illustrated in **Fig. 2.**

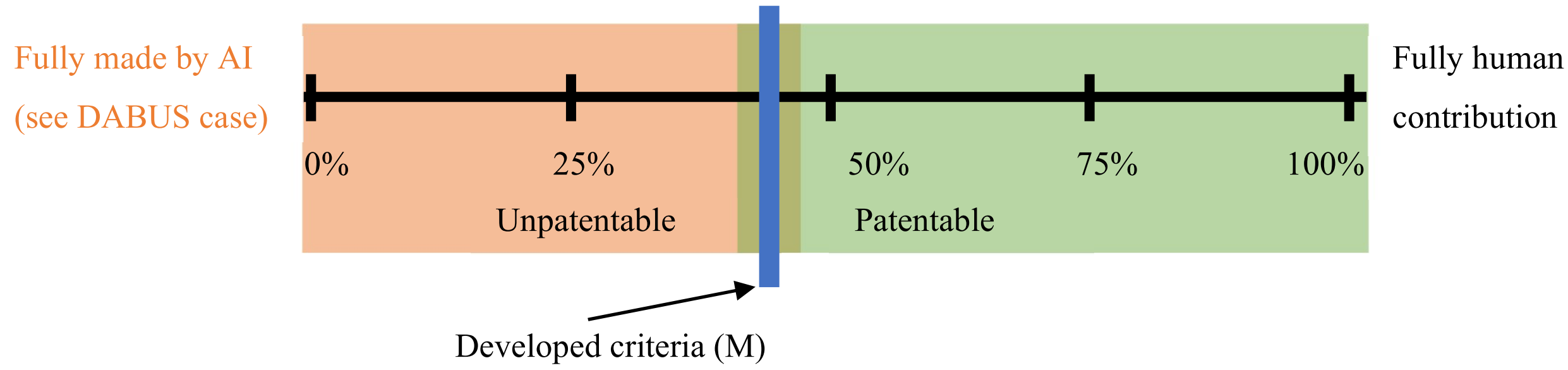

***Fig. 2** Strict approach in creation test cases for 5 cases with human contribution from 0% to 100% (black lines)*

Therefore, such an experiment allows determination of the exact value of *M*, which in turn determines the minimum value of human contribution in AI-assisted inventions necessary and sufficient for such an invention not to be rejected by the patent office due to the lack of human contribution. It is worth noting that in practice the application of the strict approach will differ significantly, but the proposed methodology provides the first steps for a discussion formalizing towards the idea proposed in the research.

### *3.3. Broad approach in creating test cases*

In contrast or even in addition to the mentioned strict approach in creation test cases*,* current study also proposes broad approach in creation test cases*,* where cases are chosen both in different areas of technology and in different types of inventive process in which AI tools were involved. This approach may provide case law as an outcome according to different types of inventive process at the same time in different types of inventive process compared to strict approach, but perhaps with less measurement accuracy of human-AI contributions.

In a broad approach, the selection of case studies for conducting the proposed experiments may vary. It is necessary to consider a range of parameters that reflect the current practices in the creation of AI-assisted inventions. These parameters include the diversity of inventive processes, the number of individuals involved in the inventive process, their respective roles (such as programming, brainstorming, management, experimentation, etc.), the specific AI tools employed and their modes of application, and the domain of the final invention (such as software, medicine, mechanics, etc.).

Current state of the art regarding which AI tools and systems are currently used in inventive processes and how (different research and development processes) is quite incomplete and unsystematic. Thus, the purpose of this section is also to create a list of most used AI approaches in an inventive process to suggest further possible cases to include in our experiment.

In the report (WIPO, 2024), WIPO made a well-structured current state of the art of generative AI used in patents and scientific publications as well as emphasized key application areas for generative AI patents, that includes "s*oftware, life sciences, document management and publishing, business solutions, industry and manufacturing, transportation, security, telecommunications"*. For current research the most interesting application areas are life sciences and industry and manufacturing, as AI can be used there as assistance to produce patentable subject matter output.

For example, in life sciences, AI is already used in the drug development process and designing new molecules, that are patentable subject matters. E.g. patent application (Patent No. 17/497,190 (patent application), 2021) describes a "*system and method for generating customizable molecular structures for drug discovery*". The disclosed system receives from the user "*input data relating to small molecules; wherein the input data comprises data relating to desirable molecules and undesirable molecules*" and generate as an output "*molecular structures in accordance with the objective function of the generative model*", which can be patented itself.

Another patent application (US Patent No. 11,288,414, 2022) describes "*systems, methods, and apparatus for artificial intelligence-based manufacturing part design*". Thus, mentioned invention enables to assist human in inventing new parts design with optimal characteristics, where user is only prompting AI with "*a desired part design, objectives for the desired part design, weightings for the objectives, and similarity bounds*".

Considering current literature in the area of AI and IP (e.g., (Cautela, Mortati, Dell'Era, & Gastaldi, 2019), examples provided by USPTO together with guidance in (USPTO, 2024a)), report (WIPO, 2024), this list may include the following:

1. Prompting generative AI such as ChatGPT iteratively to create patentable output.

2. Using *in silico* simulation or generative AI in drugs development.
3. Simulation techniques or using AI tools in assisting testing (chemistry, physics, mechanics, engineering, etc.).
   E.g. in the book (Bogni & Galli, 2020) authors mentioned a case, related to a design of toothbrush, that was created with the use of neural networks. Two neural networks were trained on datasets of existing design of toothbrushes to create optimal design.
4. Intelligent data processing.
5. Image processing.

The list mentioned is a proposal and can be expanded and changed over time depending on the development in the field of AI and its usage in inventive processes.

### *3.4. Advantages and limitations of proposed approaches*

As a result of conducting the strict approach, society will develop more certain criteria of how to measure human-AI contributions in the inventive process at least for chosen and neighboring technological areas. A limitation of the strict approach is the need to conduct a separate experiment for each area of technology.

Conducting the broad approach will lead to obtaining fewer sharp criteria compared to the strict approach, but at the same time these criteria will be obtained for a wide range of areas of technology and different types of inventive process involved.

Thus, the most comprehensive approach will be a hybrid approach, when conducting both strict and broad approaches. Moreover, it will allow us to compare results of several experiments, and if the obtained results of several approaches are similar, it will confirm scientific grounds and robustness of proposed approach and chosen cases, while if results are different, it will give feedback on how to improve conducted approaches. It will also allow us to compare developed criteria for a different field of technology and inventive process, thus confirming the robustness of the proposed approach.

## 4. Experiments

To confirm the scientific grounds in the first approximation of the proposed methodology, we conducted a pilot survey between August 19 and 31, 2024 among young professionals in IP, professors in IP, and AI experts. Participants were asked to provide their opinions on whether a human in each of several hypothetical AI-assisted inventions (cases) we created met the legal criteria to be named as the inventor or patent protection should be refused. The pilot survey included twenty-three respondents, offering a diversity of perspectives across relevant expertise areas. While this pilot survey may not lend itself to broad statistical generalization, it serves as proof of concept and demonstrates the applicability of the methodology described in this paper.

The need to conduct this pilot survey arose to test our methodology in a short time. This could be the first stage of a "pilot test" for our methodology in the survey, which can be scaled up. After that, it can be used by applying patent applications to the patent offices or to the specialized third-party entity specially created to develop criteria of how to address inventorship issue in AI-assisted inventions. This entity may then submit the recommendations it develops to the legislature or the patent office. In fact, if step 2 of the proposed approach involves having inventions reviewed by a specialized third-party entity (rather than the patent office itself), the process essentially functions as a large-scale survey.

The necessity of testing our methodology through a pilot survey is further supported by other studies in the field of AI and Law, which have likewise employed survey-based approaches. In (Harasta, Novotná, & Savelka, 2024) authors discussed topic of legal documents automatically generated using AI tools. This study used a survey between professionals and law students to explore the perception legal documents based on their assumed origin (human vs. AI-generated), revealing a clear preference for human-crafted texts despite expectations of widespread future use of generative AI. At the same time, their findings underscore a prospective trend toward the automatic generation of legal documents.

Further support for the need for a pilot survey comes from a consultation with society (UKIPO, 2021a) regarding how patent system should deal with AI-devised (AI-assisted) inventions that were discussed earlier.

Thus, these two mechanisms — pilot survey (or on a larger scale — reviewing inventions by special third-party entity) and testing a proposed approach by applying patent applications can provide further policy context by testing how actual patent examination results contrast with the perceptions of those in the industry in terms of what constitutes human contribution. At the same time, it is not just about "*human contribution*" in the abstract sense, but how human contribution maps onto the legal interpretation of the specific conditions outlines in the USPTO/EPO or Patent Laws in other jurisdictions.

Returning to the pilot survey conducted, below are described six artificial cases we created using a hybrid approach. Created cases include possible patent claims together with descriptions of inventive process.

We created three examples using a broad approach, involving various types of human-AI interactions during the inventive process. Each of the three examples contains two cases with different levels of human-AI contributions, consistent with the strict approach. Detailed examples and cases for a survey are provided in Appendix A. The summary of these examples and cases is described below in table 2.

*Table 2 Summarized description of examples and cases for a survey*

| Example | Case | Contribution of AI tools | Human contribution |
|---|---|---|---|
| Example 1. Use of generative AI in creating folding bicycle. | 1 | LLM: Generate a patent (from scratch) | General Request (Idea) |
| | 2 | LLM: Edit a generated patent | Specific edits on request |
| Example 2. Use of AI that simulates the work process of existing invention with different parameters to improve it. | 3 | AI: Find optimal parameters | Setting parameters and ranges |
| | 4 | AI: Find optimal parameters | Selecting a solution based on additional criteria |
| Example 3. Use of generative AI in improving existing patent that describes method and a computer system for gestures for touch sensitive input devices. | 5 | LLM: Improve of an existing patent | General request for improvement |
| | 6 | LLM: Edit an improvement | Specific edits on request |

Summarizing, in Example 1, we made several prompts to the generative AI tool (ChatGPT 4o) to create a folding bicycle that meets specific criteria, such as being as compact as possible when folded. After the first prompt we obtained the first invention (CASE1), and after a few additional prompts we developed CASE2.

Example 2 involves using AI to simulate the work process of an existing invention with different parameters to improve it. We considered an existing patent (patent (RU Patent No. 2747882, 2021), that describes an air disinfection device. We assumed that a researcher could use an AI tool to simulate the invention's work process to increase the efficiency of air purification and disinfection by varying input parameters mentioned in the independent claim. The researcher could obtain new parameter values that differ from those claimed in the existing patent, resulting in a new technical effect (e.g., a 10% increase in air purification and disinfection efficiency). The improved claim could thus be new, non-obvious, and patentable. Moreover, it would not infringe on the existing patent claim, as it uses different, non-claimed parameter values. Example 2 contains two inventions, CASE3 and CASE4, with additional human contribution.

Example 3 focuses on using generative AI to improve an existing patent. We used ChatGPT 4o to prompt improvements to a well-known and frequently cited patent assigned to Apple Inc. (US Patent No. 8,479,122, 2013), which describes a method and a computer system for gestures on touch-sensitive input devices. We asked ChatGPT 4o to enhance the existing patent in a way that would not infringe the original patent claim while improving how gestures are performed on touch-sensitive devices. Example 3 includes two inventions, CASE5 and CASE6, with additional human contribution.

The survey questions are described in Appendix B. The questions primarily address whether the person should be named as an inventor in each of the six cases and inquire about the respondent's experience in patent law and AI.

Detailed survey results are provided in Appendix C and summarized in **Fig. 3** below. The answer options to the questions were selected in accordance with a Likert scale. Twenty-three people participated in the survey.

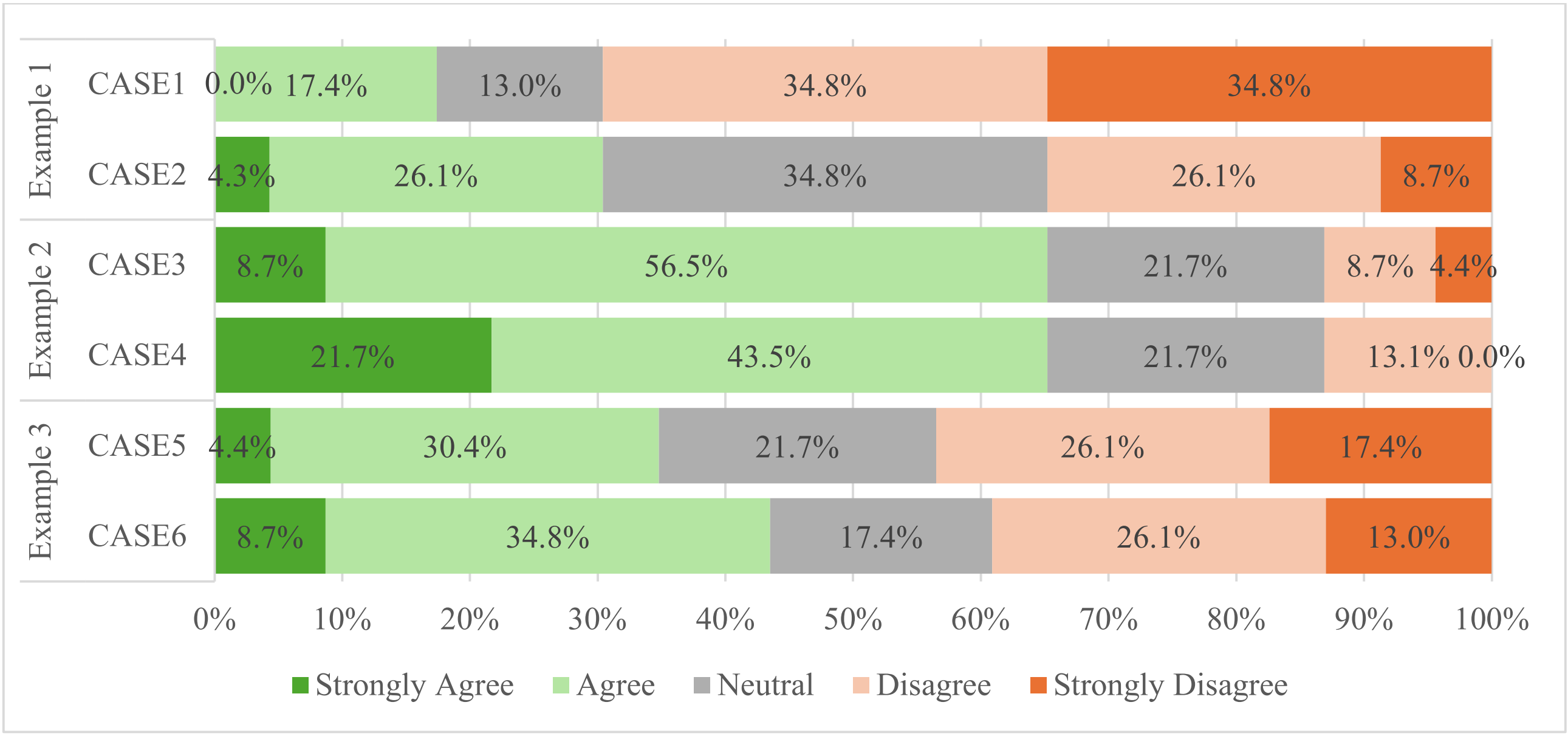

***Fig. 3*** *Survey results summary*

According to CASE1 (in example 1), most respondents (69%) disagreed that the human researcher met the legal criteria to be named as inventor, while only 17% agreed. Adding an additional step of human contribution (an additional prompt to ChatGPT) reduced the percentage of those who disagreed to 35%, while other 35% respondents answered neutral and 30% agreed.

In both CASE3 and CASE4, 65% of respondents agreed or strongly agreed that the human researcher meets the legal criteria to be named as the inventor, 13% disagreed or strongly disagreed and 22% answered "neutral". The difference is that in CASE4, more respondents selected "strongly agree" rather than just "agree". Thus, results for example 2 show that use of AI to simulate the work process of an existing technology by choosing input parameters by a person increases the number of respondents who agree that a person's contribution is sufficient to be recognized as the inventor of such AI-assisted invention opposed to LLM prompting in example 1.

In CASE5, 35% agreed, while 44% disagreed, whereas in CASE6, 44% agreed, and 39,1% disagreed. That is respondents' opinion indicates greater confidence in the need to recognize a person as an inventor in the case of using LLM to improve an existing invention in example 3 than to create an invention from scratch using LLM in an example 1. Although this confidence is still smaller for both types of usage of LLM (example 1, 3) by humans than for using AI tools for simulation in example 2.

Thus, the pilot survey results show that incorporating an additional step of human contribution/involvement in the inventive process leads to greater consensus that the human researcher meets the legal criteria to be named as the inventor. Moreover, simply prompting generative AI is generally considered insufficient for the human researcher to be recognized as the inventor, whereas utilizing simulation AI tools to enhance existing technologies is viewed as sufficiently contributive for the human researcher to be considered the inventor of the improved invention.

Even based on the results of the pilot survey conducted, it is already possible to draw some conclusions and recommendations for patent offices. The greatest progress in this matter has been achieved by the approach of the USPTO in their initial "Inventorship Guidance for AI-Assisted Inventions" (USPTO, 2024a). They proposed the consideration of two artificial examples of AI-assisted inventions (one is related to the use of a LLM model and another with the use *in silico* methods in drug discovery) and how patent examiners should consider them in the question of human inventorship depending on human contribution, see (USPTO, 2024b). If the approach reflected in the initial guidance (USPTO, 2024a) is taken as a basis, our pilot survey and the cases proposed in it indicate potential value in further expanding the guidance through additional examples involving different categories of AI tools (i.e., not only LLM and *in silico*). Such examples may include AI-assisted inventions created with the use of AI tools to simulate work processes to increase efficiency or other characteristics of existing inventions (like in our proposed example 2). In general, the more examples future guidance will include such that are considered from different areas of technology, with different uses of AI-tools (and creative techniques), and with a variety of human contributions, the better for a more complete understanding of how to measure human contribution in AI-assisted inventions for patent applicants. This, in essence, encapsulates the core idea of our hybrid approach and particularly in our strict and broad approaches.

As for other patent offices, including EPO and JPO, given the development of the inventorship issue in this paper, the proposed methodology in our study and the pilot survey results, we can recommend starting from releasing guidance for examination with a large and varied number of examples of AI-assisted inventions and how to address inventorship issue in each of them, where these AI-assisted inventions should cover different areas of technology, different uses of AI-tools, and with a variety of human contributions. The first edition of the guidance could include several of the most popular areas of technology among applicants, as well as the techniques and AI-tools used to create AI-assisted inventions. New examples may be added as the inventorship issue develops in future editions of inventorship guidance.

It is worth noting that such inventorship guidance may be a good guide to consider the inventorship issue in AI-assisted inventions at the beginning. But already now there is a need to make changes or clarifications directly to the Patent Laws of how to address inventorship issue in AI-assisted inventions. This is where conducting an experiment in accordance with the methodology we propose in our study can help.

Given the current lack of case law on inventorship issue in AI-assisted inventions, the results of the pilot survey provide valuable preliminary insights. These findings can serve as a foundation for developing test patent applications that explore various scenarios of AI involvement and human contribution. Such test cases may help stimulate legal discourse and judicial consideration, ultimately contributing to the establishment of case law that addresses the novel challenges posed by AI-assisted inventions.

During the creation of test cases and the analysis of the pilot survey results, it is also become evident that a new challenge is emerging: in the near future, there could be a significant increase in AI-assisted patent applications that represent incremental improvements on existing patents. AI tools may also be employed to improve existing patents while avoiding infringement. Consequently, patent law will need to address these challenges in the patent granting process and provide guidance for examiners to maintain a balance between social welfare, technological progress, and the rights of patentees. Stakeholders should also consider these issues in their business processes, for instance, by developing internal guidelines for the use of AI tools in the inventive process, as well as in patent enforcement and portfolio management.

After conducting our pilot survey (August 19-31, 2024) and finalizing the main analysis, we identified an article (Henry H. Perritt, 2025), in which the author describes filing a patent application for an AI-assisted invention: on November 13, 2024, the author asked ChatGPT to design "*a generative AI system to search for and identify products and services that potentially infringe on a patent*". Perritt drafted the entire specification with AI assistance and filed the patent application with the USPTO together with a declaration describing how AI was used. In the declaration, he also stated that he believes that the initially AI-generated application exceeds USPTO's Example 1, Scenario 3 (USPTO, 2024b), according to which additional prompting of the AI system and other steps such as conducting experiments and changes to the invention's elements are sufficient to recognize a human as an inventor. The final rejection issued on March 17, 2025, did not include any objection related to inventorship. However, the application was rejected based on 35 U.S.C. § 101 subject matter eligibility and 35 U.S.C. § 103 obviousness grounds.

One of the possible explanations for the absence of an objection or any feedback related to inventorship in Perritt's case is the possible revision of USPTO guidance (USPTO, 2025a) at the time of the final rejection of his invention, following U.S. policy changes introduced on January 23, 2025 (USPTO, 2025b), which directed agencies to review and revise policies in order to reduce barriers to AI innovation. Another explanation is that Perritt's contribution to the AI-assisted invention was considered sufficient under the initial USPTO guidance, highlighting the importance of such guidelines, which provide inventors with a clearer understanding of what constitutes sufficient human contribution and what does not.

Such strong interest from academia and legislators, together with the rapid pace of AI development, may lead to another revision of laws and guidelines, further reinforcing the relevance of the inventorship issue addressed in this study. The results of our pilot survey, as well as the Perritt case as an illustrative example, provide preliminary support for the proposed methodology and indicate the potential value of scaling the experiment to further examine legal uncertainty in AI-assisted inventions.

However, it is stressed that the analysis in this paper focusses on the interpretation of human inventorship as a criterion for patentability. The widespread deployment of AI-technology will also affect other substantive patent criteria, such as novelty and most notably inventive-step/non-obviousness (Abbott, 2019). Determining the minimum threshold for human inventorship contribution is independent of an application's evaluation in relations to these other substantive criteria. However, when implementing a pilot survey in accordance with the methodology outlined in this paper, it needs to be clear to respondents that the focus is on the issue of machine/human inventorship contribution, which should not be conflated with the concept of inventive step and/or obviousness.

## 5. Conclusions and Discussion

This study addressed an emerging topic at the intersection of areas of AI and IP: the nature and contribution of using AI tools in AI-assisted inventions and whether such contributions mean a human cannot be considered the inventor. The research summarized the current legal landscape and case laws regarding inventorship and its application to AI-assisted inventions in most jurisdictions. The current work also identified an inventorship issue in AI-assisted inventions, particularly the lack of case law addressing this issue and the absence of developed criteria for measuring human-AI contributions in defining inventorship.

As a novel approach, this research proposes discussing a condition under which a new and bold experiment can be implemented to create case law and resolve the legal uncertainty regarding inventorship in AI-assisted inventions. This experiment could be initiated by society, involving stakeholders specializing in AI. The proposed approach essentially includes steps such as creating inventions (cases) with varying levels of human-AI contributions or types of AI involvement in the inventive process, applying for a patent with an explanation of the inventive process and AI involvement, and receiving either a patent grant or refusal. The results of such decisions may be used in the future in determining criteria of which amount of human-AI contributions is enough to obtain a patent. The research proposed a methodology for conducting this experiment and selecting cases that best reflect the current state of AI use in the inventive process. Utilizing this approach will help identify the best methods for measuring human contributions to AI-assisted inventions and determining inventorship. The proposed approach can be also used in other fields of IP law, where legal uncertainties exist, especially in relation to new emerging technologies in Patent Law and Copyright Law.

To validate the scientific basis of the proposed methodology, we conducted a pilot survey among young professionals in IP, professors, and AI experts, asking them to provide their opinions on whether inventorship should be granted or patent protection refused for each of six proposed cases. The survey results indicated that incorporating an additional step of human involvement in the inventive process leads to a greater consensus that the human researcher meets the legal criteria for being named as the inventor. In addition, the use of different inventive processes is evaluated differently. For example, using simulation AI tools versus generative AI tools is more often considered in favor of recognizing a human as the inventor of the AI-assisted invention.

Based on pilot survey results we propose several recommendations for different patent offices of how guidance for examinations can already be improved in considering inventorship issue in AI-assisted inventions. The main recommendation is to release or improve guidance for examination with a large and varied number of examples of AI-assisted inventions and how to address inventorship issue in each of them, where these AI-assisted inventions should cover different areas of technology, different uses of AI-tools, and with a variety of human contributions.

Our methodology and pilot survey provide a basis for developing test patent applications that explore varying levels of AI involvement and human contribution. These cases could stimulate legal discourse and judicial consideration and help shape case law on AI-assisted inventions. Our results can also inform policy by comparing actual patent examination outcomes with industry perceptions of what constitutes human contribution. With this proof of concept, this mechanism based on a survey methodology can be scaled up to generate robust results that can inform patent examination practices.

**Appendix A. Case examples for a survey**

**Example 1.** Use of generative AI in creating folding bicycle. Example 1 contains CASE1 and CASE2 with different human-AI involvement.

Below there are several prompts made by researcher to ChatGPT 4o that show the process of creation an invention in which both human (researcher) and AI-tool (ChatGPT) were involved.

*Prompt 1: Make a patent claims for a folding bicycle that folds as compactly as possible.*

*ChatGPT: Creating patent claims for a folding bicycle requires a careful and detailed description of the invention, highlighting its unique features and advantages. Here's a set of sample patent claims for a folding bicycle that folds as compactly as possible:*

**Patent Claims (CASE 1)**

**1.** A folding bicycle comprising:

- a frame including a first frame portion and a second frame portion,
- a hinge mechanism connecting said first frame portion to said second frame portion, enabling said first and second frame portions to pivot relative to each other between an extended position for riding and a folded position for compact storage,
- a locking mechanism to secure said first and second frame portions in said extended position,
- a front wheel assembly and a rear wheel assembly attached to said frame,
- a handlebar assembly connected to said front wheel assembly and configured to pivot and fold parallel to said frame,
- a seat post assembly with a seat, wherein said seat post assembly is telescopically adjustable and configured to retract into said frame,
- wherein said front wheel assembly and said rear wheel assembly are configured to fold towards each other to minimize the overall volume of the bicycle in said folded position,
- wherein said bicycle, in said folded position, forms a compact structure with said frame portions, handlebar assembly, and seat post assembly substantially aligned in a parallel arrangement.

The researcher continued to improve the resulting Patent Claims (CASE1) by creating additional prompts to ChatGPT 4o.

*Prompt 2: Change the claim for frame that contains one frame portion*

*[The intermediate ChatGPT answer was hidden as insignificant.]*

*Prompt 3: Change a claim so that bicycle contains single frame, but folding fork and folding feathers*

*ChatGPT: Certainly! Here is a revised set of patent claims for a folding bicycle that includes a single frame, a folding fork, and folding feathers (chainstays):*

**Patent Claims (CASE2)**

**1.** A folding bicycle comprising:

- a single continuous frame,
- a folding fork assembly attached to a front end of said frame,
- folding chainstays attached to a rear end of said frame,
- a hinge mechanism integrated into said folding fork assembly and said folding chainstays, enabling the bicycle to fold between an extended position for riding and a folded position for compact storage,
- a locking mechanism to secure said folding fork assembly and said folding chainstays in said extended position,
- a front wheel assembly attached to said folding fork assembly,
- a rear wheel assembly attached to said folding chainstays,
- a handlebar assembly connected to said folding fork assembly and configured to pivot and fold parallel to said frame,

- a seat post assembly with a seat, wherein said seat post assembly is telescopically adjustable and configured to retract into said frame,
- wherein said folding fork assembly and said folding chainstays are configured to fold towards the central portion of said frame to minimize the overall volume of the bicycle in said folded position,
- wherein said bicycle, in said folded position, forms a compact structure with said frame, folding fork assembly, folding chainstays, handlebar assembly, and seat post assembly substantially aligned in a parallel arrangement.

**Example 2.** Use of AI that simulates the work process of existing invention with different parameters to improve it. Example 2 contains CASE3 and CASE4 with different human-AI involvement.

There is an existing patent (RU Patent No. 2747882, 2021), that disclosure the following invention:

Claim 1. A device for air disinfection, comprising a housing that includes an air intake chamber with an inlet window, an air discharge chamber with an outlet window, equipped with a fan, an irradiation chamber located in front of the air discharge chamber, containing a source of ultraviolet radiation, characterized in that between the chamber air intake and irradiation chamber there is an air purification chamber, and a carbon filter is installed at the entrance to the air purification chamber, and a double partition is made at the exit from the air purification chamber with a distance of 5 to 10 mm at the hole, in each of which there is a filter in the form of copper wires installed mutually perpendicular and connected to an electric current source, and the inlet window of the air intake chamber is equipped with a fan, and the inner surface of the irradiation chamber is made of reflective material.

The researcher used an already existing AI tool to simulate work process of mentioned invention to increase the efficiency of air purification and disinfection varying input parameters of the invention. As an input parameters he chose followings:

*Parameter a*. Distance at the hole of the air purification chamber (in existing invention it is from 5 to 10 mm. In simulation *parameter a* varies from 0 mm to 100 mm.

*Parameter b*. Type of the partition at the exit from the air purification chamber which is made. In the existing invention there is a double partition, but in simulation *parameter b* takes values from a set B = {single, double, triple, …}.

*Parameter c.* Type of the material of the inner surface of the irradiation chamber. In the existing invention it is made from reflective material. In simulation *parameter c* takes values from a set C = {reflective, non-reflective, semi-reflective}.

*Parameter d*. The existing invention contains fan 1 in an outlet window, and fan 2 in the inlet window. In simulation *parameter d* characterizes the presence and combination of fans and takes values from a set D = {fan 1 and fan 2, only fan 1, only fan 2, no fans}.

*Parameter e.* The existing invention contains filter in the form of copper wires installed mutually perpendicular. In simulation *parameter e* determines type of the shape of the filter and takes values from a set E = {perpendicular, honeycomb shape, with angle of 45˚, …}.

After simulations made by AI tool, suppose that the best parameters are the following: *a* = 4 mm, *c* = reflective, *d* = fan 1 and fan 2, *e* = honeycomb shape. Suppose that new values, the efficiency of air purification and disinfection increased by 10% compared to the existing invention.

Thus, as values of the parameters *a* (4mm) and *e* (honeycomb shape) are different from values of these parameters in the original invention and as new technical result is previously unknown (increasement of the efficiency of air purification and disinfection by 10% comparing to the original patent), this improvement of the existing invention can be considered as new and non-obvious — see CASE3 below.

**New claim (CASE 3):**

Claim 1. A device for air disinfection (the same as in original claim), characterized in that distance at the hole of the air purification chamber is 4 mm, filter is in the form of honeycomb shape.

After that researcher also realized that according to simulation results using only fan 1 (*parameter d*) together with air purification chamber is *a* = 4 mm, filter is in the form of honeycomb shape the efficiency of air purification and disinfection is increased by 8% comparing to the original patent according to researcher's calculations, and only 2% less comparing to CASE3.

Thus, as values of the parameters *a* (4mm) and *e* (honeycomb shape) and *d* (only fan 1) are different from values of these parameters in the original invention and as new technical result is previously unknown (increasement of the efficiency of air purification and disinfection by 8% and decreasing energy consumption by 10% comparing to the original patent), this improvement of the existing invention can be considered as new and non-obvious — see CASE4 below.

**CASE4:**

Claim 1. A device for air disinfection (the same as in original claim), characterized in that distance at the hole of the air purification chamber is 4 mm, filter is in the form of honeycomb shape, only fan 1.

**Example 3.** Use of generative AI in improving existing patent. Example 3 contains CASE5 and CASE6 with different human-AI involvement.

There is well-known frequently cited (more than 3000 forward citations) patent, assigned to Apple Inc, (US Patent No. 8,479,122, 2013), that describes method and a computer system for gestures for touch sensitive input devices.

Below there are several prompts made by a researcher to ChatGPT 4o that show the process of creation new invention by improving an existing patented invention US8479122B2 in which both human (researcher) and AI-tool (ChatGPT) were involved.

***Prompt 1:*** Improve the following patent claim in such a way that improved claim will not infringe the following claim and will improve the way gestures are performed on touch sensitive devices: What is claimed is: 1. A method for manipulating graphical user interface (GUI) objects displayed on a touch sensing device, comprising: displaying a plurality of GUI objects on a multipoint touch screen of the touch sensing device, the multipoint touch screen providing a primary display for the touch sensing device; detecting one or more concurrent touches at different points on the multipoint touch screen, directly on or in proximity to the plurality of the displayed GUI objects; associating the one or more touches to a first GUI object from the displayed plurality of GUI objects in response to a relevance ranking of touch data associated with the one or more touches to each of the GUI objects; maintaining the association of the one or more touches with the first GUI object despite a presence of a second GUI object having a relevance ranking that exceeds the relevance ranking of the first GUI object if a difference between the relevance rankings of the second GUI object and the first GUI object is less than a predetermined threshold amount; and interpreting the multiple touches as a single gesture for performing a particular operation on the associated GUI object, the operation determined based on the associated GUI object.

***<u>ChatGPT:</u>*** To improve the method of performing gestures on touch-sensitive devices while ensuring that the new claim does not infringe the existing one, the revised claim will focus on a novel approach to gesture detection and interpretation. This approach will be distinct in how it handles multiple touches, associations, and gestures while enhancing user interaction.

Here's an improved patent claim:

**Claim 1 (CASE5)**: A method for manipulating graphical user interface (GUI) objects displayed on a touch sensing device, comprising:

- displaying a plurality of GUI objects on a multipoint touch screen of the touch sensing device, the multipoint touch screen providing a primary display for the touch sensing device;
- detecting one or more concurrent touches at different points on the multipoint touch screen, directly on or in proximity to the plurality of the displayed GUI objects;
- dynamically determining a gesture zone based on the location and movement of the detected touches, wherein the gesture zone encompasses one or more GUI objects and adjusts in real-time as touch points are added, removed, or moved;
- prioritizing the GUI objects within the gesture zone based on a combination of touch intensity, duration, and proximity to the center of the gesture zone;
- associating the touch data with the GUI object having the highest priority within the gesture zone, wherein the association is updated dynamically as touch points change;
- performing a gesture interpretation process by analyzing the movement trajectory, speed, and sequence of the touch points within the gesture zone, and determining a corresponding operation based on a predefined set of gesture rules;
- executing the determined operation on the associated GUI object, wherein the operation includes at least one of resizing, rotating, translating, or triggering a contextual action specific to the associated GUI object.

**Explanation and Improvements:**

1. **Gesture Zone Determination**: The improved claim introduces the concept of a "gesture zone" that is dynamically determined and adjusted based on the touch points' location, movement, intensity, and duration. This is a novel approach that focuses on the interaction area rather than individual touch points.
2. **Prioritization Mechanism**: The claim specifies a method for prioritizing GUI objects within the gesture zone based on various factors, including touch intensity, duration, and proximity to the gesture zone's center. This prioritization approach is different from a relevance ranking based on static criteria.

3. **Dynamic Association**: The association of touch data to a GUI object is dynamically updated as the gesture zone changes, which is different from maintaining an association despite higher relevance rankings of other objects.

4. **Enhanced Gesture Interpretation**: The method includes a detailed process for gesture interpretation that considers movement trajectory, speed, and touch sequence within the gesture zone. This interpretation method is designed to improve accuracy and flexibility in gesture-based operations.

5. **Non-Infringement**: The new claim avoids infringement by not using the specific relevance ranking and threshold-based association approach described in the original claim. Instead, it introduces an entirely different method for determining how gestures are recognized and applied to GUI objects.

This improved claim enhances the user experience on touch-sensitive devices by making gesture recognition more intuitive, responsive, and capable of handling complex interactions without infringing on the existing patent claim.

The researcher continued to improve the resulting Patent Claims (CASE5) by creating additional prompts to ChatGPT 4o.

***Prompt 2.*** Change the term “the highest priority” so that it can vary in certain range of values. Make following claims as an open list claims: “Prioritizing the GUI objects within the gesture zone based on a combination of touch intensity, duration, and proximity to the center of the gesture zone”, “performing a gesture interpretation process by analyzing the movement trajectory, speed, and sequence of the touch points within the gesture zone”.

***ChatGPT:*** Certainly! Here’s the updated patent claim with the requested changes:

**Claim 1 (CASE6)**: A method for manipulating graphical user interface (GUI) objects displayed on a touch sensing device, comprising:

- displaying a plurality of GUI objects on a multipoint touch screen of the touch sensing device, the multipoint touch screen providing a primary display for the touch sensing device;
- detecting one or more concurrent touches at different points on the multipoint touch screen, directly on or in proximity to the plurality of the displayed GUI objects;
- dynamically determining a gesture zone based on the location and movement of the detected touches, wherein the gesture zone encompasses one or more GUI objects and adjusts in real-time as touch points are added, removed, or moved;
- prioritizing the GUI objects within the gesture zone based on a combination of touch intensity, duration, and proximity to the center of the gesture zone, wherein the priority of a GUI object is assigned a value within a defined range and is dynamically adjustable based on changes in the gesture zone and associated touch data;
- associating the touch data with the GUI object having a priority value within the defined range, wherein the association is updated dynamically as touch points change;
- performing a gesture interpretation process by analyzing, among other factors, the movement trajectory, speed, and sequence of the touch points within the gesture zone; and
- executing the determined operation on the associated GUI object, wherein the operation includes at least one of resizing, rotating, translating, or triggering a contextual action specific to the associated GUI object.

**Explanation of Changes:**

1. **Varying Priority Value**:
   - The term “the highest priority” has been changed to indicate that the priority of a GUI object can vary within a defined range. This allows for more flexibility in determining which GUI object should be associated with the touch data, rather than always selecting the one with the absolute highest priority.

2. **Open-List Claims:**
   - The claims “Prioritizing the GUI objects within the gesture zone based on a combination of touch intensity, duration, and proximity to the center of the gesture zone” and “performing a gesture interpretation process by analyzing the movement trajectory, speed, and sequence of the touch points within the gesture zone” have been structured as open-list claims by adding the phrase “among other factors.” This allows for additional criteria or factors to be considered without being explicitly listed, thereby broadening the scope of the claims.

These modifications provide greater flexibility and scope to the claims, ensuring that the invention can adapt to different scenarios and use cases while maintaining legal defensibility.

**Appendix B. Survey questions**

Below is a description and questions for respondents attached to the pilot survey conducted.

Please answer the following questions regarding your qualifications and experience in the fields of intellectual property (IP) and artificial intelligence (AI).

1. What is your experience in the field of IP?
   a. (Basic) Basic understanding or no experience.
   b. (Beginner) Beginner specialist (student studying IP).
   c. (Expert) Experienced patent engineer, patent attorney, professor in Patent Law.
2. What is your experience in the field of AI?
   a. (Basic) Basic understanding or no experience.
   b. (Beginner) Beginner specialist (student studying AI, machine learning).
   c. (Expert) AI expert, professor in AI.

Please provide feedback data (optional):

3. Name of respondent.
4. Current position held and organization.
5. Email address for obtaining feedback.

Six patent claims (cases) for AI-assisted inventions are provided below. Each case was made with the involvement of both human (researcher) and AI tools. For each case there is a description of the inventive process, that is, how exactly a human used AI tools to produce a final result (patent claim) that can be patented. It is assumed that all patentability criteria are met (novelty, inventive step, industrial applicability).

For each case, respondents will be asked the following question: considering the human contribution and the extent of AI tool usage in creating the invention as described in the claim, do you believe the human researcher meets the legal criteria to be named as the inventor? We also would greatly appreciate your specific comments.

6. The human researcher meets the legal criteria to be named as inventor.
   a. Strongly Disagree
   b. Disagree
   c. Neutral
   d. Agree
   e. Strongly Agree
7. Please briefly explain your choice (optional).

**Appendix C. Survey results**

The pilot survey was conducted between August 19 and 31, 2024. It was sent to LL.M. in IP students, professors in the field of IP, patent attorneys, and experts in the field of AI. They were asked to provide their opinions on whether the human meets the legal criteria to be named as the inventor, or patent protection should be refused in each of six AI-assisted inventions (cases) we created. Overall, twenty-two people participated in the survey, which can be considered a representative sample. The answer options to the questions were selected in accordance with a Likert scale.

Below in tables T1-T5 there are the pilot survey results with comments provided by respondents. There are twenty three people who took part in the survey with different experiences in IP and AI.

*Table T3 Responses regarding experience in the fields of IP and AI*

| What is your experience in the field of IP? | | What is your experience in the field of AI? | |
|---|---|---|---|
| Basic understanding or no experience | 21.80% | Basic understanding or no experience | 47.80% |
| Beginner specialist (student studying IP) | 30.40% | Beginner specialist (student studying AI, machine learning) | 34.80% |
| Experienced patent engineer, patent attorney, professor in Patent Law | 47.80% | AI expert, professor in AI | 17.40% |

*Table T4 Responses for CASES 1-6*

| **Examples** | **Example 1** | | **Example 2** | | **Example 3** | |
|---|---|---|---|---|---|---|
| **Cases** | | | | | | |
| **Question: The human researcher meets the legal criteria to be named as inventor?** | **CASE1** | **CASE2** | **CASE3** | **CASE4** | **CASE5** | **CASE6** |
| Strongly Disagree | 34.8% | 8.7% | 4.4% | 0.0% | 17.4% | 13.0% |
| Disagree | 34.8% | 26.1% | 8.7% | 13.1% | 26.1% | 26.1% |
| Neutral | 13.0% | 34.8% | 21.7% | 21.7% | 21.7% | 17.4% |
| Agree | 17.4% | 26.1% | 56.5% | 43.5% | 30.4% | 34.8% |
| Strongly Agree | 0.0% | 4.3% | 8.7% | 21.7% | 4.4% | 8.7% |

*Table T5 Respondent's comments for CASE1 and CASE2*

| Experience in IP | Experience in AI | CASE1 | Comments for CASE1 | CASE2 | Comments for CASE2 |
|---|---|---|---|---|---|
| Expert | Basic | Strongly Disagree | Human only provided for the disired and common functional features they wished the invention to have, not providing any input concerning how the function would be achieved. | Strongly Disagree | Same comment as before |
| Beginner | Beginner | Strongly Disagree | Nothing is invented, just described how does it work | Disagree | Very high level description without any specific |
| Basic | Basic | Disagree | Effort of creating such a patent is quite insignificant, in case if we allow such patent submissions to be recognized it might be exploited by patent trolls to generate tons of junk patents that will have a devastating impact on tech industry. | Neutral | In this case the researches not just generated patent with AI, but started fine-tuning it using generative AI as tool. It still might be questionable though if the effort is of enough significance to consider human researcher to meet the the legal criteria. |
| Expert | Beginner | Strongly Disagree | The prompt given by the human was too simple, and so their contribution is not substantial. I cannot see how this prompt would be linked to some 'inventive spark' in which the human was (partially) responsible for the non-obvious step. | Disagree | The human contribution was more here and shows an understanding of the technical subject matter. However, I do not find that the adjustment to the prompts reflects that the human input is connected to the innovative concepts in the claims. |
| Expert | Basic | Strongly Disagree | With respect to example 1 (1/2), the human researcher does not meet the legal criteria to be named as an inventor (who contributed to the patentable invention) because the prompt 1 contains only general requirements for the claim. The claim limitations generated by the AI tool go far beyond the specification of the request. Thus, human involvement in the resulted claim is minimized. | Strongly Disagree | With respect to example 1 (2/2), the human researcher does not meet the legal criteria to be named as an inventor (who contributed to the patentable invention) because subsequent prompts contain only a slight specification for several features of the previously generated claim. The AI tool generated new additional features though, that were not even requested. Thus, human involvement in the resulted claim is still very limited. |
| Expert | Expert | Strongly Disagree | As I understand the case, the human researcher did not make any invention, did not conceive it, did not contribute to the putting into practice of the invention, etc. These are all fundamental criteria to be considered an inventor, and as there are not being fulfilled (as the human researcher did not conceive the invention as claimed), the human researcher is not the invention, nor is the AI system | Disagree | I still think that the abovementioned fundamental criteria have not been fulfilled, hence the human researcher cannot be an inventor, nor is the AI system |
| Basic | Expert | Disagree | Lack of new ideas in the prompt | Neutral | Prompt isn't specific enough |
| Basic | Beginner | Disagree | repetitive language used, not too descriptive language | Agree | more appropriate and advanced language |
| Basic | Expert | Agree | It does look that AI did a major contribution to the proposed invention. However, the request, the initial idea and the final validation of the answer is performed by a human. | Strongly Agree | After the first request, the human did not just agree with invention but requested further improvements until answer is fully validated. |
| Expert | Basic | Neutral | I gave a neutral answer, because for a more precise answer I do not have enough information about what has already been done by the researcher. Simply getting help from AI to write a patent formula is one story, where the researcher has already come up with something and is the author. Unwinding a chain of design features of a particular object using AI, and then trying to patent it using AI is another story, where all the actions of the researcher were built only on the competent use of prompts, i.e. in this approach he should not be the author. | Disagree | What is described here looks more like a selection of solutions. Therefore, in my opinion, the authorship here may already be in doubt. |
| Beginner | Basic | Strongly Disagree | The human researcher's input into the machine merely contains the requirement of a bike which folds as compactly as possible. | Disagree | Though the human researcher has given more intellectual input (a bike consisting of a single frame but folding fork and folding feathers), in my opinion it still is not enough to be named as the inventor. Most of the technical features can still be contributed to the AI. |
| Expert | Beginner | Strongly Disagree | No inventive contribution (problem is known; prompt does not contribute anyhow to | Agree | Solution to the technical problem provided by user of AI, whereas AI limits itself to compile a |

| | | | the solution described in the claim). | | description of the resulting bike by adding obvious details. |
|---|---|---|---|---|---|

*Table T6 Respondent's comments for CASE3 and CASE4*

| Experience in IP | Experience in AI | CASE3 | Comments for CASE3 | CASE4 | Comments for CASE4 |
|---|---|---|---|---|---|
| Expert | Basic | Agree | Any other software simulation tools could have been used. AI was not decisive. Use of simulation tools to improve invention is a common practice. Thus, I don't see why it should not be allowed to put the human as the inventor. Having said that, patentability criteria could be a problem, not being as easily met as described for this exercise. | Agree | Same comment |
| Beginner | Beginner | Agree | AI is used as tool (program) to find optimal parameters | Agree | Same as case 3 |
| Basic | Basic | Neutral | Even though the case is more complicated then simple use of generative AI and fine-tuning patent by tweaking the prompts, still it's quite questionable on significance of human researcher effort | Agree | In comparison to Case 3, Case 4 seems like the one that involves human researcher analysis and expertise to accomplish the invention and therefore meets the legal criteria to be names as inventor |
| Expert | Beginner | Agree | Here I can see the human contribution as being linked to the innovation in the claims. My instinct that this should not be patented is more about non-obviousness though (as if the AI was able to optimize these parameters, then maybe it was obvious to the PHOSITA). But I do think that the human contribution is there in a significant way. If a lead engineer had given the optimisation problem to her research team, then I would see all of the parties as co-inventors. Here, the fact that the research team is changed for AI does not change the fact that the lead engineer meets the standard of (co)inventor. | Agree | Same as before! |
| Expert | Basic | Strongly Agree | With respect to example 2 (1/2), the human researcher may meet the legal criteria to be named as an inventor (who contributed to the patentable invention) because he has invented new different possible improvements of the known invention. The AI tool was used to simulate (i.e. to | Strongly Agree | With respect to example 2 (2/2), the human researcher meets the legal criteria to be named as an inventor because he has performed additional scrutiny based on previous results of AI tool simulation resulting in |

| | | | test) different combination of parameters in order to achieve some better efficiency, but not to invent them. Moreover, AI tool used known ways of how these parameters should have been combined to be simulated, as they were described in the existing patent. The resulted claim contains the features, that were invented by the human researcher and tested by the AI tool. | | another improvement of the known invention. |
|---|---|---|---|---|---|
| Expert | Expert | Agree | Conception and putting into practice has been done by the human researcher. The AI is merely used as a tool, very much like any inventor would use any other computer program. | Agree | same reason as previous case |
| Basic | Expert | Agree | | Strongl y Agree | Important that a human performs own final assessment, while AI output is only used as in input to human's final assessment. |
| Expert | Basic | Neutral | I disagree with his authorship more than I agree. But there is a nuance. The thing is that the researcher entered some data into the model that he had originally invented, for example, honeycomb shape. That is, he already had some ideas about alternatives to the invention, but did not understand whether they made sense. The model only calculated all his alternatives and gave a result on efficiency. | Neutral | Here is the same story as in the example below. It seems that initially the researcher generated alternatives (which allows us to consider him the author), and then verified their effectiveness. |
| Expert | Beginner | Agree | AI only performs experiments that were conceived and prepared by a human inventor. Situation does not appear significantly different from what happens in any laboratory, where technicians perform experiments often conceived by others. | Agree | A fortiori, considering my previous answer. |

*Table T7 Respondent's comments for CASE5 and CASE6*

| Experience in IP | Experience in AI | CASE5 | Comments for CASE5 | CASE6 | Comments for CASE6 |
|---|---|---|---|---|---|
| Expert | Basic | Strongly Disagree | Inventor merely defined the desired benefits without provding any creative input as to how the technical features coul de achieved | Strongly Disagree | Same comment |
| Beginner | Beginner | Neutral | | Neutral | This sounds more like idea but not invention |
| Basic | Basic | Disagree | The same thoughts as for the Case 1 | Disagree | Seems like it's not advanced much in comparison to Case 5 or Case 1, I'd disagree that human researcher meets the legal criteria. |
| Expert | Beginner | Agree | I feel like the inventive spark was undertaken by the human inventor here. What the AI did was just help in claim drafting, or rather ensure that the drafting is narrow enough the ensure novelty. | Agree | Again, I see the human as being responsible for the invention, and the AI helping in claim drafting. In this case, the human first asks the AI to narrow the claims, then asks it to broaden it a bit. This seems to be a matter of first ensuring novelty and then aiming for a widened scope of protection. These are separate issues from the concept of inventorship. |
| Expert | Basic | Strongly Disagree | This example 3 resembles the example 1, wherein the human researcher asks the AI tool not to create an invention from scratch, but to modify existing claim to generate a new one. The human input is not only insufficient, but also refers to a request unrelated to a patentable subject matter (improve the claim so as not to infringe the existing claim). | Disagree | The comment is similar to that of example 1. However, this time it looks like the AI tool has only supplemented the claim with corrections offered by the human researcher. If the AI tool is used as an instrument for formulating ideas that have been invented by humans, then there should not be an inventorship issue. |
| Expert | Expert | Disagree | Also here, the human researcher did not conceive the invention, but merely asked an AI system to write a new patent claim. | Agree | This is probably a bit of a difficult one to decide upon. The human researcher added specific parameters for the invention to be claimed. In that respect, the human researcher could be seen as having contributed to conception of the invention, one of the fundamental requirements to be considered an inventor |
| Basic | Expert | Disagree | Contribution of the human researcher is not significant | Disagree | |
| Expert | Basic | Strongly Disagree | I don't see any contribution of the researcher to the solution here. Only the work of the model. | Strongly Disagree | For me, minor clarifications from the researcher did not change the fact that the solution was generated by the model. |
| Expert | Beginner | Disagree | No relevant contribution to the development of the allegedly inventive solution: all that the human does is saying "this is the state of the art; try to improve it". | Disagree | Human contribution does not relate to the invention but to the way in which it is claimed. |